\documentclass[conference,comsoc]{IEEEtran}
\usepackage{cite}

\ifCLASSINFOpdf 
\usepackage[pdftex]{graphicx}
\else
\usepackage[dvips]{graphicx}
\fi

\usepackage[cmex10]{amsmath}
\usepackage{algorithmic}
\usepackage[ruled,norelsize]{algorithm2e}

\usepackage{array}

\usepackage{mdwmath}
\usepackage{mdwtab}
\usepackage{todonotes}

\usepackage{eqparbox}

\ifCLASSOPTIONcompsoc
\usepackage[caption=false,font=normalsize,labelfont=sf,textfont=sf]{subfig}
\else
\usepackage[caption=false,font=footnotesize]{subfig}
\fi

\usepackage{fixltx2e}
\usepackage{url}
\usepackage{amsfonts}
\usepackage{amssymb}
\usepackage{amsthm}
\usepackage{bm}
\usepackage[ruled]{algorithm2e}
\usepackage[utf8]{inputenc}
\usepackage{booktabs}
\usepackage{url}
\usepackage{cite}
\usepackage{flushend}
\usepackage{epstopdf}

\newcounter{tempEquationCounter} 
\newcounter{thisEquationNumber}

\begin{document}
\title{{Satellite Direct-to-Device from Low Earth Orbit: Techno-Economic Analysis of a Global Non-Terrestrial Network}}

\author{\IEEEauthorblockN{Adnan Aijaz, Peizheng Li, Sajida Gufran}
\IEEEauthorblockA{Bristol Research and Innovation Laboratory, Toshiba Europe Ltd., U.K.\\
Email: \{firstname.lastname\}@toshiba-bril.com}}
\markboth{IEEE XYZ -- Submitted for Publication}%
{Shell \MakeLowercase{\textit{et al.}}: Bare Demo of IEEEtran.cls for Journals}

\maketitle
\begin{abstract}
\boldmath
Low Earth orbit (LEO) satellites and satellite direct-to-device (D2D) technology are at the heart of the next-generation global connectivity which promises direct access to space-based broadband services for unmodified 3GPP-compliant handsets.  With a rapidly evolving ecosystem, it is important to evaluate the feasibility, cost-effectiveness, and profitability of these services. By assessing the technological aspects as well as economic implications, stakeholders can make informed decisions about investment, development, and deployment strategies. 
This paper presents a comprehensive techno-economic analysis (TEA) framework for evaluating LEO-based satellite D2D systems. The framework integrates a global satellite constellation model, radio propagation aspects including atmospheric and rainfall attenuation models compliant with ITU-R recommendations, 3GPP-compliant capacity calculations, realistic global population data, and an all-encompassing cost model accounting for both capital and operational expenses associated with space and ground segments. Further, the framework evaluates three different architectural options for realizing a global non-terrestrial network (NTN) for satellite D2D services. With an emphasis on reproducibility, the framework has been implemented through significant enhancements to an open-source tool. The economic assessment reveals that global satellite D2D services can be provided at a monthly cost per subscriber which is comparable to terrestrial services while achieving a positive return on investment (ROI). Moreover, the results show the potential of Open RAN technology for realizing cost-effective satellite D2D services.  
\end{abstract}


\IEEEpeerreviewmaketitle

\section{Introduction}
\IEEEPARstart{L}{ow Earth orbit} (LEO) satellites are emerging as a new frontier for global connectivity transformation. Orbiting at altitudes of 150 to 2000 km, LEO satellites potentially offer low-latency, high-speed Internet access. LEO satellites are envisaged as a key component of the non-terrestrial networks (NTNs) for beyond 5G and 6G systems~\cite{wang2025non}. The global LEO satellite market\footnote{{https://satellitemarkets.com/market-trends/global-leo-satellite-market-forecast-2024-2030}} is expected to reach 34B USD by 2030, growing at a CAGR of 13.28\% from 2024.

Satellite direct-to-device (D2D) technology allows standard/unmodified user equipment like handsets to directly communicate with satellites. Driven by technological advancements, evolving regulatory frameworks, and market demands, the satellite D2D market is projected to experience significant growth over the next few years. It represents a paradigm shift and a significant step forward in seamless global connectivity. Ongoing activities in 3GPP Releases 17, 18, and 19 are paving the way for standardized D2D communications~\cite{lin20215g, lin2025bridge}. Regulatory frameworks for satellite D2D services are also evolving in different parts of the world.

Techno-economic analysis (TEA) is a tool/technique that integrates technical and economic assessments to evaluate the viability of a technology. 
Combining technical feasibility, cost analysis, and market potential, it provides a comprehensive understanding of the impact and value proposition of a technology. In context of satellite D2D, TEA can help assess the technical performance, cost-effectiveness, and market potential of satellite-based communication for broadband applications. It can also identify key factors influencing the success of these systems, such as regulatory frameworks, market demand, and technological advances.

\subsection{Related Work}
Some recent studies have investigated techno-economic aspects of satellite-based communication systems. Osoro and Oughton~\cite{TEA_RW1} proposed an open-source engineering-economic framework to assess LEO constellations such as Starlink, OneWeb, and Kuiper. Their findings indicate that LEO networks can provide competitive capacity (e.g., $\sim$25 Mbps per user), but only in extremely sparse areas ($\leq$0.1 users/km$^2$). 
Li \textit{et al.}~\cite{TEA_RW2} developed a simulation-based model for evaluating the cost-efficiency of LEO mega-constellations in the context of 6G. 
They show that optimizing RF characteristics can reduce user-centric costs.
Toka \textit{et al.}~\cite{TEA_RW3} extended the TEA concept to integrated airspace and non-terrestrial networks combining LEO, HAPS, and UAVs. Their results suggest that a layered architecture enables flexible and cost-effective 6G deployment across diverse geographies.
Chiha \textit{et al.}~\cite{chiha2020techno} introduced the concept of a third-party broker to manage resource allocation between mobile network operators (MNOs) and satellite network operators (SNOs) in integrated satellite-5G systems. The broker improves efficiency and cost-effectiveness by abstracting satellite resources and streamlining inter-operator negotiations. 
Ye \emph{et al.}~\cite{ye2024techno} examined multi-satellite collaboration through high-rate optical inter-satellite links (OISLs). Their study found that OISLs can triple throughput and reduce transmission costs by up to 70\% compared to RF-based solutions.

\subsection{Contributions and Outline}


To this end, this work introduces a TEA framework for satellite D2D systems, grounded in a global NTN based on a LEO satellite constellation. In contrast to prior studies that primarily address link-level performance, inter-satellite cooperation, or broad business considerations, our TEA adopts a comprehensive end-to-end methodology that closely integrates technical system design with detailed economic evaluation.

The framework is developed to assess the technical performance, cost-effectiveness, and market potential of satellite D2D services. 
The primary objective of this work is to offer a holistic view of the techno-economic dimensions of satellite D2D by integrating system-level design considerations, cost data for both space and ground segments, population density estimates, and other relevant real-world factors. The main contributions are summarized as follows 
\begin{itemize}
\item We have developed a comprehensive TEA framework for satellite D2D by integrating satellite constellation modeling, radio propagation characteristics, 3GPP-compliant capacity estimation, global population data, and detailed cost structures for both space and ground segments.

\item To ensure reproducibility, the TEA framework has been implemented by significantly enhancing a general-purpose TEA tool for satellite communication systems. The tool incorporates three distinct architectural designs for realizing a global NTN.

\item We conducted a techno-economic assessment of two representative real-world systems with distinct design and operational characteristics. For each system, we analyzed user capacity performance, monthly subscription cost, and return on investment (ROI) across three architectural models: bent-pipe, regenerative, and 3D Open RAN. 

\end{itemize}

The rest of the paper is organized as follows. Section \ref{sect_tea} describes the proposed TEA framework. Results for performance evaluation and techno-economic insights are provided in Section \ref{sect_perf}. The paper is concluded in Section \ref{sect_cr}.

\section{Proposed TEA Framework}\label{sect_tea}

\subsection{Methodology and Approach}
\begin{figure}[tp]  
\centering
\includegraphics[width=0.98\linewidth]{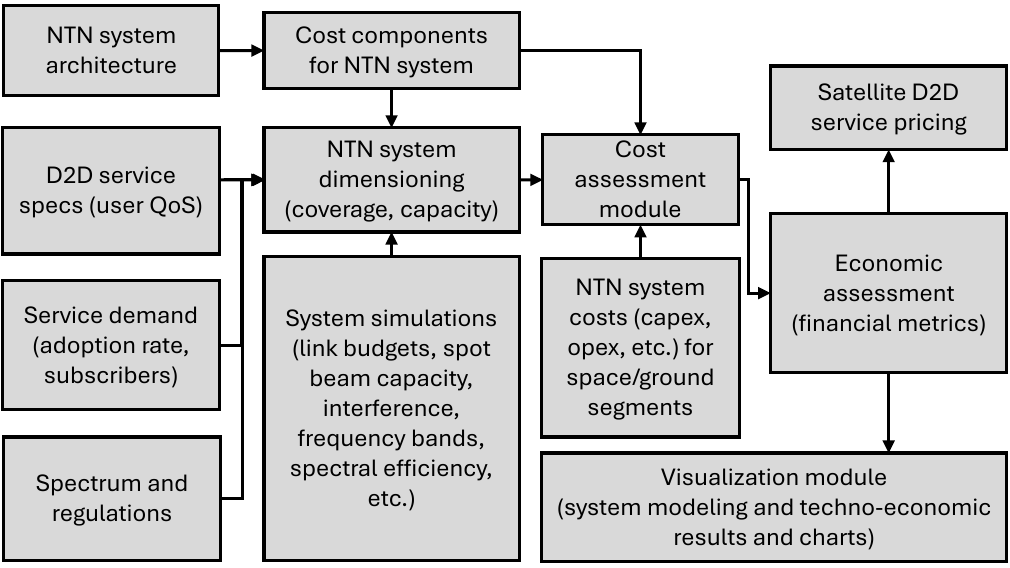}
\caption{Overview of the TEA methodology and approach used in this work.}
\label{fig:methodlology}
\vspace{-1em}
\end{figure}

The methodology underlying the TEA framework is illustrated in Fig. \ref{fig:methodlology}, combining comprehensive technical modeling with economic evaluation. The framework is structured around dimensioning a global NTN system to support satellite D2D coverage and capacity. This dimensioning process is guided by service-level specifications such as required quality-of-service (QoS), projected D2D demand, and regulatory considerations (e.g., spectrum availability). It is further supported by system-level simulations accounting for link budgets, spot beam configurations, and interference management. Architectural elements are translated into cost components across the space and ground segments.

The cost structure is detailed through capital expenditure (Capex) and operational expenditure (Opex) components, which feed into a cost assessment function. This function estimates key financial indicators including total cost of ownership (TCO), net present value (NPV) of future expenditures, and return on investment (ROI). These metrics inform service pricing strategies and revenue projections. The framework outputs include quantitative results—such as performance and cost charts—as well as qualitative insights into broader socioeconomic impacts.

\begin{figure}
\centering
\includegraphics[scale=0.6]{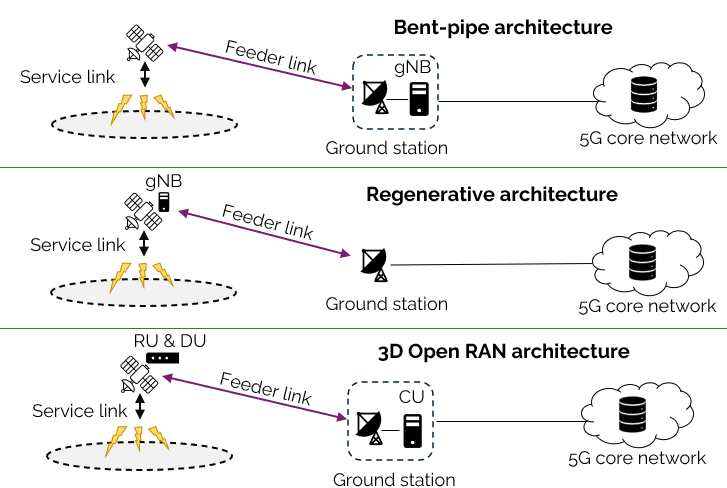}
\caption{NTN architecture options considered for techno-economic analysis.}
\vspace{-1.5em}
\label{sys_arch}

\end{figure}

\subsection{System Architecture}

Our TEA is centred around three key architectural options for satellite-based communication systems which are illustrated in Fig.~\ref{sys_arch} and described as follows\footnote{A single satellite is shown in the architecture instead of the constellation for simplicity.}.

\textit{Bent-pipe} -- In this architecture (also known as the transparent architecture), the satellite simply acts as a repeater in the sky whereas the base station is located on the ground.

\textit{Regenerative} -- In this architecture, the satellite acts as a base station in the sky, i.e., the base station is onboard the satellite platform. 

\textit{3D Open RAN} -- This architecture exploits the RAN functional splits such that some components of the base station are onboard the satellite platform while other components are located on the ground.

The key components of the system architecture that influence the TEA are described as follows. 

\textit{Service link} – The service link provides coverage in the illuminated area on the ground. 

\textit{Feedber link} – The feeder link (or the backhaul link) connects the satellite to the ground station. 

\textit{gNB} – The gNB refers to the 5G base station which can be monolithic or disaggregated in architecture. 

\textit{Ground station} – The ground station hosts necessary infrastructure for integration of non-terrestrial segment into the terrestrial segment of the network. 

\textit{5G core network} – The 5G core network is responsible for managing a wide range of network functions including authentication, security, session and mobility management, traffic/data management, and routing.


\subsection{Satellite Constellation Model}
We consider a global satellite constellation model that consists of multiple  LEO satellites. The constellation is designed to provide global coverage with minimal latency and high throughput. The key parameters of the constellation model include the number of satellites, their orbital altitude, inclination, and the frequency bands used for communication. The model also considers the inter-satellite links (ISLs) for efficient data routing and communication between satellites.
The key parameters of the constellation model include the following:

\begin{itemize}
  \item \textit{Number of satellites} -- The number of satellites in the constellation is determined by the desired coverage area and the  orbital altitude of the satellite. \textcolor{black}{The number of satellites should be chosen to provide global coverage with a minimum data rate guarantee.}
  \item \textit{Orbital altitude} -- The orbital altitude determines the distance between the satellite and the earth. \textcolor{black}{While operation at higher altitudes provides better coverage, it can also lead to increased interference and limited user performance. }
  \item \textit{Inclination} -- The inclination of the satellite's orbit determines the coverage area and the number of satellites required to provide global coverage. A higher inclination provides better coverage in polar regions.
  \item \textit{Frequency bands} -- The  choice of frequency bands is influenced by available bandwidth, regulatory constraints, interference, and propagation characteristics.
  \item \textit{ISLs} -- The ISLs are used for inter-satellite data routing and communication. The ISLs should be designed to minimize latency and maximize throughput for inter-satellite communication in regnerative architectures. 
  \item \textit{Interference management} -- The model considers interference management techniques to minimize the impact of interference on the satellite D2D system.
\end{itemize}


\subsection{Satellite Orbit and Location Distribution}

\begin{figure}
\centering
\includegraphics[scale=0.4]{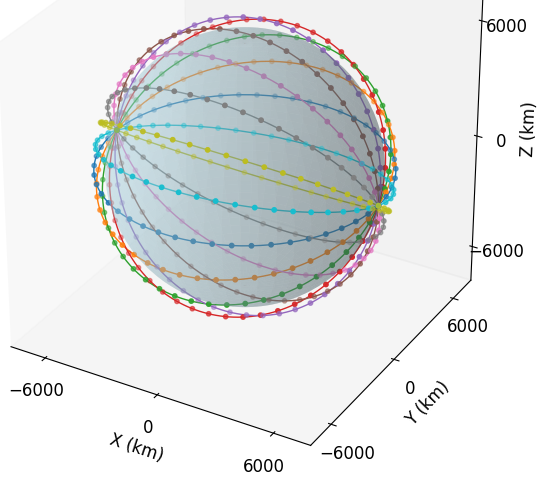}
\caption{LEO satellites distribution with 10 orbits and 50 satellites per orbit.}
\label{Sat_orbits}
\vspace{-1.5em}
\end{figure}

We adopt a uniform distribution for modeling the satellite locations in orbit as showing in Fig.~\ref{Sat_orbits}. This simplified spatial model assumes that satellites are evenly distributed along their orbital planes and across the constellation, facilitating tractable system-level analysis of coverage and capacity. While more complex stochastic models such as the Cox distribution~\cite{choi2024cox} are used to capture spatial heterogeneity in user or device locations, a uniform satellite distribution is sufficient and appropriate for TEA as it  represents well-designed LEO constellations with planned, deterministic deployment patterns.

\subsection{Radio Propagation Model}
The radio propagation model is a key component of the TEA as it determines the coverage area, signal quality, and capacity of the satellite D2D system. The model takes into account various factors such as path loss, atmospheric attenuation, rainfall attenuation, and user density. It is designed to provide accurate estimates of the signal strength and quality at different distances and frequencies, allowing for effective planning and optimization of the satellite D2D system. The propagation model is based on the ITU-R P.618-13 recommendation for satellite systems~\cite{ITU-R_P618-13}. The model accounts for free space path loss, atmospheric absorption, and rain attenuation. 
The path loss is calculated as 
\begin{equation}
\textrm{PL} = 20\log_{10}(d) + 20\log_{10}(f) + K,
\end{equation}
where \(d\) is the distance in km, \(f\) is the frequency in MHz, and \(K\) is a constant.  The effective isotropic radiated power (EIRP) is calculated as: 
\(\text{EIRP} = \text{tx\_power} \cdot \text{ant\_gain}\),
where tx\_power is transmit power and ant\_gain is transmit antenna gain.

\subsubsection{Rainfall Attenuation}
Rainfall attenuation is a significant factor in satellite communication, especially in tropical and subtropical regions. The model uses the ITU-R P.618-13 recommendation to estimate the rainfall attenuation based on the rain rate, frequency, and polarization of the signal. 
Based on~\cite{ITU-R_P618-13}, the rainfall attenuation is calculated as
\(A_r = k \cdot R^{\gamma}\),
where \(A_r\) is the rainfall attenuation in dB, \(k\) is the specific attenuation coefficient, \(R\) is the rain rate in mm/h, and \(\gamma\) is the frequency-dependent exponent. The specific attenuation coefficient and the exponent are determined based on empirical data and models provided by ITU-R.

\subsubsection{Atmospheric Attenuation}
Atmospheric attenuation is a critical factor in satellite communication, particularly at higher frequencies. The model incorporates the ITU-R P.676-12 recommendation~\cite{ITU-RP676} to estimate the atmospheric attenuation based on the frequency, elevation angle, and atmospheric conditions. It considers factors such as oxygen and water vapor absorption, which can significantly affect signal quality.

\subsection{Interference Model}
The interference model is based on uniformly distributed satellites in circular orbits with inclined configurations. Within each orbit, satellites are evenly spaced based on arc-length to maintain consistent angular separation. Multiple orbits are generated by incrementally rotating each orbital plane by a fixed 
inclination angle.

A ground receiver, positioned 1 meter above the Earth’s surface, connects to the nearest satellite at its lowest orbital point. Other satellites with clear line-of-sight, determined via a dot-product method to account for Earth’s curvature, are treated as interference sources.

From the visible satellites, the five closest to the receiver are identified as dominant interferers. For each, the signal’s off-axis angle is calculated, and the antenna gain is modeled with a cosine roll-off plus minor suppression outside the main lobe. Interference power is computed via free-space path loss and summed to obtain total inter-satellite interference.

Intra-satellite inter-beam interference is modeled by spacing beams across a 90° coverage sector, with each beam having a 90° width. Off-axis gains from non-primary beams follow a cosine profile, and their interference contributions are summed to determine total inter-beam interference.

The total interference power at the receiver combines both inter-satellite and intra-beam components, capturing spatial and angular interference characteristics in LEO constellations while preserving analytical simplicity through streamlined gain models.
Based on this model, the inter-satellite interference is calculated by
\begin{equation}
\text{Int}_\text{S} = \sum_{i=1}^I\frac{\text{EIRP}_i \cdot \text{G}(\theta_i)}{\text{PL}_i},
\end{equation}
where \(I\) is the total number of neighboring interfering satellites, 
and \(\text{G}(\theta_i)\) is off-axis antenna gain, which is calculated using a cosine profile based on the angular position of the interfering beam as \( \text{G}(\theta_i) = \cos^2(\theta_i) \).
Likewise, inter-beam interference, excluding the serving beam, is calculated as
\begin{equation}
\text{Int}_\text{B} = \sum_{i=2}^B\frac{\text{EIRP}_i \cdot \text{G}(\theta_i)}{\text{PL}_i},
\end{equation}
where \(B\) is the total number of beams per satellite. The total interference power \(\text{Int}_\text{T} \) is calculated as $\text{Int}_\text{T} = \text{Int}_\text{S} + \text{Int}_\text{B}$.

\subsection{User Density and Traffic Model} 
\begin{figure}[tp]  
\centering
\includegraphics[width=0.98\linewidth]{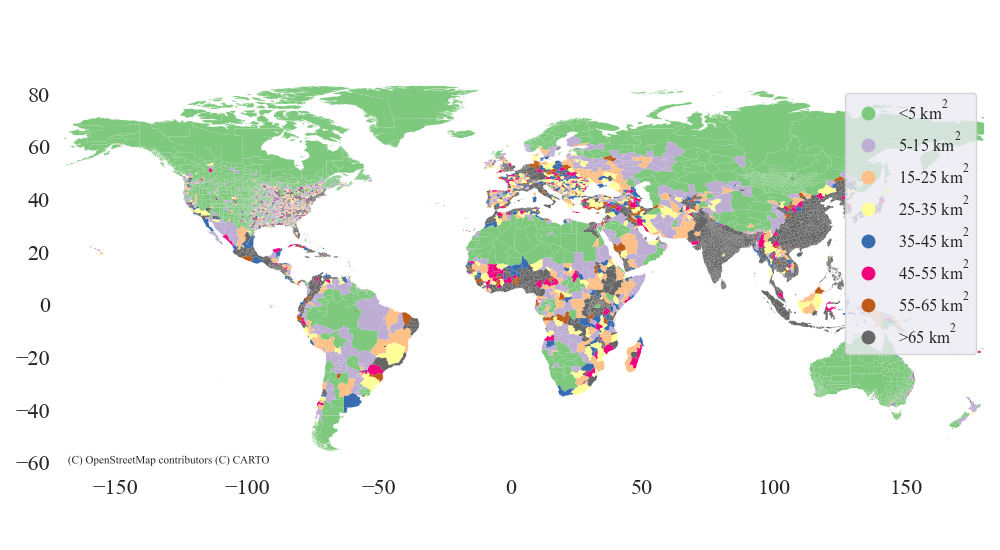}
\caption{User density distribution across geographical regions.}
\label{fig:user_density_distribution}
\vspace{-1.5em}
\end{figure}

We obtain the user density (users per square kilometer) data based on the population density in different regions from WorldPop \cite{worldpop2021}, and a visualized representation of the user density distribution is shown in Fig.~\ref{fig:user_density_distribution}. The figure illustrates the variation of user density across different geographical regions, highlighting areas with high user concentration and potential demand for satellite D2D services. Each country is divided into multiple regions based on official administrative boundaries, such as provinces and districts.

User density and traffic patterns are modeled from expected demand and service needs, factoring in user counts, average data rates, adoption rates, overbooking factors, and peak-hour conditions. The overbooking factor (OBF) represents active users during peak periods, while the adoption rate reflects the proportion of the population using NTN services. This model simulates realistic temporal traffic variations, enabling accurate capacity and coverage analysis.
Using OBF and adoption rate, we calculate active users density per km\textsuperscript{2} as below.
\begin{equation}
\text{Active}_{\text{users}} = \frac{\text{Average}_{\text{density}} \cdot \text{Adop}_{\text{rate}}}{\text{OBF}}
\label{active_users}
\end{equation}

\subsection{Coverage and Capacity Calculations}
The coverage and capacity calculations are performed using the radio propagation model, interference model and the traffic model. The coverage area is determined by the signal strength at different distances from the satellite, while the capacity is calculated based on the available bandwidth, user density, and average data rate per user. The model provides a comprehensive analysis of the coverage and capacity of the satellite D2D system, allowing for effective planning and optimization of the network. First, we calculate the signal-to-interference-plus-noise ratio (SINR) at the receiver using transmit EIRP, path loss, noise, interference, rain attenuation, and atmospheric losses. Then, we derive the spectral efficiency (SE) based on 3GPP-based modulation and SE mapping tables~\cite{3gpp_ts_38_214_v15_9_0}. 
Once we have the SE, we compute the single-beam capacity as 
\(\text{Beam}_{\text{cap}} = \text{SE}  \cdot  \text{DL}_{\text{bw}}\),
where \(\text{DL}_{\text{bw}}\) is downlink bandwidth of a beam. 
Using the  capacity of the beam, we calculate satellite capacity as 
\(\text{Sat}_{\text{cap}} = \text{Num}_{\text{beams}}  \cdot  \text{Beam}_{\text{cap}}\),
where \(\text{Num}_{\text{beams}}\)  is the total number of beams per satellite.
The total capacity per km\textsuperscript{2} is given by
\begin{equation}
\text{Capacity}_{\text{km\textsuperscript{2}}} = \frac{\text{Sat}_{\text{cap}}}   {\text{Sat}_{\text{area}}},
\label{capacity_km2}
\end{equation}
where \({\text{Sat}_{\text{area}}}\) is the single satellite footprint area on Earth's surface.
Finally, we get per user capacity  as follows. 
\begin{equation}
\text{User}_{\text{capacity}} = \frac{\text{Capacity}_{\text{km\textsuperscript{2}}}}   {\text{Active}_{\text{users}}}
\end{equation}

\subsection{Cost Model}

The cost model is designed to estimate the Capex and Opex associated with the deployment and operation of the satellite D2D system. It includes various cost components such as satellite manufacturing, launch costs, ground station infrastructure, and ongoing operational costs. The cost estimates used for the TEA are based on publicly available figures, industry benchmarks, and cost estimation to provide a comprehensive financial analysis of the system.
The breakdown of the cost model is as follows.
\begin{itemize}
    \item \textit{Capex} - This includes one-time costs associated with the deployment of the satellite D2D system, such as satellite manufacturing, launch costs, ground station infrastructure, and regulatory fees.
    \item \textit{Opex} - This includes ongoing operational costs such as maintenance, in-orbit replenishment, staffing, research and development, and marketing.

\item \textit{Net Present Value (NPV)} - The net present value is used to estimate the total cost of ownership. It is calculated over a fixed period (indicated by $Y$) using a discount rate ($r$) as follows.
\begin{equation}
    \text{NPV} = \text{Capex} + \sum_{i=0}^Y\frac{\text{Opex}}{(1+r)^i} 
\end{equation}

\end{itemize}

Using NPV, the technological and economical models are integrated to calculate the cost per subscriber, monthly subscription costs, the net revenue, and the ROI.

Some of the key components of the cost model are described as follows. 

\begin{itemize}
    \item \textit{Satellite launch cost} -- The launch cost for a satellite depends on its mass which includes the communication payload among other items. It is calculated as a product of satellite mass (in kg) and the launch cost/kg for a certain mission. It also includes a fixed cost component.  

    \item \textit{Spectrum acquisition cost} -- The spectrum acquisition cost refers to the initial cost of acquiring spectrum for service and feeder links.

    \item \textit{Spectrum Maintenance cost} -- This is the annual cost of maintaining the spectrum for service and feeder links. 

    \item \textit{Monolithic gNB cost} -- This is the cost of an integrated single vendor gNB from a supplier. It is further split into the cost of the baseband unit (BBU) and the remote radio unit (RRU). 

    \item \textit {Disaggregated gNB cost} -- This is the cost of a disaggregated gNB as per the Open RAN framework. It is further split into the software costs of the centralized unit (CU), the distributed unit (DU), the hardware cost of the CU/DU computing platforms, and the cost of the radio unit (RU). 

    \item \textit {Satellite platform cost} -- This is the cost of building a satellite platform.

\end{itemize}

\subsection{Open-source TEA Tool}
\begin{figure*}[htbp]
\centering
\includegraphics[width=0.85\linewidth]{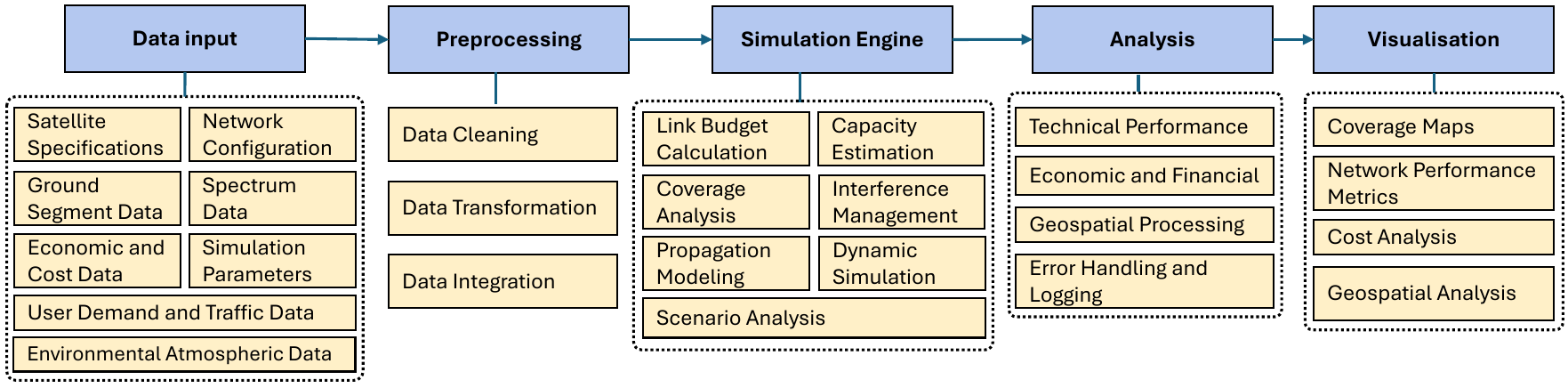}
\caption{Functionality of the enhanced open-source TEA tool for satellite D2D systems.}
\label{fig:tea_tool_functionality}
\vspace{-1.5em}
\end{figure*}
We have made significant enhancements to the open-source TEA tool developed by Osoro and Oughton~\cite{TEA_RW1} for satellite-based communication system. While the original tool focused on the techno-economic feasibility of generic LEO satellite constellations, our enhancements target satellite D2D scenarios. To summarize, we have introduced the following capabilities.
\begin{itemize}
    \item Integration of different system architectures for satellite D2D, including bent-pipe, regenerative, and 3D Open RAN configurations.
    \item Enhancements to radio propagation models, incorporating atmospheric and rainfall attenuations, and realistic link-level path loss. 
    \item Incorporation of a global satellite constellation model, with configurable orbital parameters, inclination, and satellite distribution.
    \item Enhancements to the cost modeling modules, encompassing both capex and opex components, with configurable parameters for launch, platform, infrastructure, operations, and other costs.
    \item Capability of analyzing a variety of scenarios and configurations, including satellite altitudes, orbital distributions, frequency bands, and available bandwidth.
\end{itemize}
These enhancements are crucial for satellite D2D TEA, especially for insights into coverage, capacity, and economic viability for different deployment strategies. The functional blocks of the enhanced TEA tool are shown in Fig.~\ref{fig:tea_tool_functionality}.

\section{Performance Assessment and Insights}\label{sect_perf}

\subsection{Scenarios and Parameters}
As a case study, we conduct system-level performance evaluation for a hypothetical global satellite network operator providing D2D services via a NTN. We consider two different system designs for D2D, which are referred to as System A and System B, respectively. The two systems differ in terms of design and operational aspects.

\begin{itemize}
    \item \textbf{System A} - Inspired by a Starlink-like or a Lynk-like system\footnote{Similar to SpaceX's second generation Starlink satellites (\url{https://www.starlink.com/}) or Lynk satellites (\url{https://lynk.world/}).} capable of creating dozens of spot beams. 

    \item \textbf{System B} - Inspired by a BlueWalker- or BlueBird-like system\footnote{Similar to AST SpaceMobile's BlueWalker3 or BlueBird Block 2 satellites (\url{https://ast-science.com/spacemobile-network/})} capable of creating hundreds to thousands of spot beams. 
    
\end{itemize} 

The beamforming capabilities are largely dictated by the size of the phased array on these satellites. We consider two different operating scenarios for the global NTN. 
\begin{itemize}
    \item Operation in dedicated spectrum for satellite services, i.e., the mobile satellite service (MSS) spectrum with relatively higher bandwidth availability; e.g., in the 2.7 GHz band with up to 100 MHz available bandwidth.

    \item Operation in leased spectrum from a mobile network operator (MNO) with relatively limited bandwidth availability; e.g., in the 2 GHz 5G spectrum band (i.e., band n1) with up to 10 MHz available bandwidth.
\end{itemize}

We consider an altitude of 500 km, transmit EIRP of 60 dBW, and beamwidth of 90 degrees for both systems. Our capacity calculations are based on link budget analysis and adaptive modulation and coding schemes based on link-level SINR~\cite{3gpp_ts_38_214_v15_9_0}. The path loss for 2.7 GHz and 2 GHz bands is \textcolor{black}{158.44} and \textcolor{black}{150.07} dB, respectively. We assume mmWave satellite backhaul with link capacity of 2 Gbps per link. We consider a maximum of 4 backhaul links at a single ground station (GS).  
Based on the global population density data, we consider an average user density of 181 users per km\(^{\textrm{2}}\). The default adoption rate is 2\% while the OBF is 20. 

While we investigate user performance under varying number of satellites and spot beams, our economic assessment is based on providing a minimum data rate guarantee of 10 Mbps per user while the D2D system operates in the following configurations. 

\begin{itemize}
    \item System A: 64 spot beams per satellite; 100 MHz per spot beam. 
    \item System B: 1000 or 2800 spot beams per satellite; 10 MHz per spot beam. 
\end{itemize}

These figures are realistic considering phased array size and beamforming capabilities of prominent LEO satellite systems targeting D2D services. Considering available specifications for Starlink, BlueWalker3, and BlueBird Block 2 satellites, we assume that a System A satellite weighs 800 kg whereas a System B satellite weighs 1500 kg and 5850 kg, for 1000 and 2800 spot beams, respectively.

We make the following key assumptions on system design and cost for investigating the impact of different architectural options (shown in Fig. \ref{sys_arch}). 

\begin{itemize}
    \item A satellite  with regenerative payload uses up to 4 ISLs. 

    \item Compared to bent-pipe, the regenerative architecture reduces backhaul/feeder link capacity requirements by 20\% which is realistic considering the overheads associated with gNB to core network connectivity (at the IP layer) versus relaying the air-interface data. 

    \item The cost of building an Open RAN gNB via disaggregated multi-vendor software stack and general-purpose hardware is 25\% less than that of a monolithic gNB based on proprietary hardware/software from a single vendor. This assumption is based on experience from a previous government-funded 5G project~\cite{10453148}.

    \item A single CU can support up to 16 DUs as part of the disaggregated 3D Open RAN architecture. 

    \item Compared to bent-pipe, the 3D Open RAN architecture increases the backhaul/feeder link capacity requirements by 10\% which is realistic considering overheads associated with IP layer connectivity between the gNB and core network and RAN-level connectivity between the DUs and the CUs, in case of regenerative and 3D Open RAN architectures, respectively. 

    \item A satellite with regenerative payload carries a monolithic base station (gNB) weighing  35 kg. 

    \item A satellite in 3D Open RAN architecture carries a payload of 20 kg for gNB components (RU and DU). 

\end{itemize}

The key cost parameters used for assessment are summarized in  Table~\ref{tab:cost_params}. Note that some of the parameters are for the overall constellation. 

\begin{table}[]
\caption{Key cost model parameters for the case study}
\centering
\begin{tabular}{|l|ll|}
\hline
\textbf{Parameter}           & \multicolumn{1}{l|}{\textbf{Type}} & \textbf{Cost (US Dollars; USD)} \\ \hline
Satellite launch            & \multicolumn{1}{l|}{Capex}       & 0.5M       \\ \hline
Satellite payload       & \multicolumn{1}{l|}{Capex}            & 10k/kg            \\ \hline
Monolithic gNB & \multicolumn{1}{l|}{Capex}            & 100k           \\ \hline
GS equipment        & \multicolumn{1}{l|}{Capex}            & 100k (per backhaul link)            \\ \hline
GS antenna system        & \multicolumn{1}{l|}{Capex}            & 20k (per backhaul link)            \\ \hline
Inter-satellite link (ISL)       & \multicolumn{1}{l|}{Capex}            & 25k (per link)            \\ \hline
Ground station build       & \multicolumn{1}{l|}{Capex}            & 0.5M           \\ \hline
Spectrum acquisition      & \multicolumn{1}{l|}{Capex}            & 300M           \\ \hline
Satellite platform (System A)     & \multicolumn{1}{l|}{Capex}            & 3M           \\ \hline
Satellite platform (System B)     & \multicolumn{1}{l|}{Capex}            & 10M (1000 spot beams)          \\ \hline
Satellite platform (System B)     & \multicolumn{1}{l|}{Capex}            & 15M (2800 spot beams)          \\ \hline
Satellite replenishment     & \multicolumn{1}{l|}{Opex}            & 1M per satellite         \\ \hline
Regulation fees      & \multicolumn{1}{l|}{Capex}            & 1M           \\ \hline
Digital infrastructure      & \multicolumn{1}{l|}{Capex}            & 2.5M           \\ \hline
Marketing     & \multicolumn{1}{l|}{Opex}            & 50M           \\ \hline
Staff (overall)     & \multicolumn{1}{l|}{Opex}            & 10M           \\ \hline
Research/Development (overall)     & \multicolumn{1}{l|}{Opex}            & 50M           \\ \hline
Maintenance (overall)     & \multicolumn{1}{l|}{Opex}            & 15M           \\ 
\hline
Profit margin     & \multicolumn{1}{l|}{Opex}            & 60\%         \\ 
\hline
\end{tabular}
\label{tab:cost_params}
\vspace{-1em}
\end{table}

\begin{figure}
\centering
\includegraphics[scale=0.2]{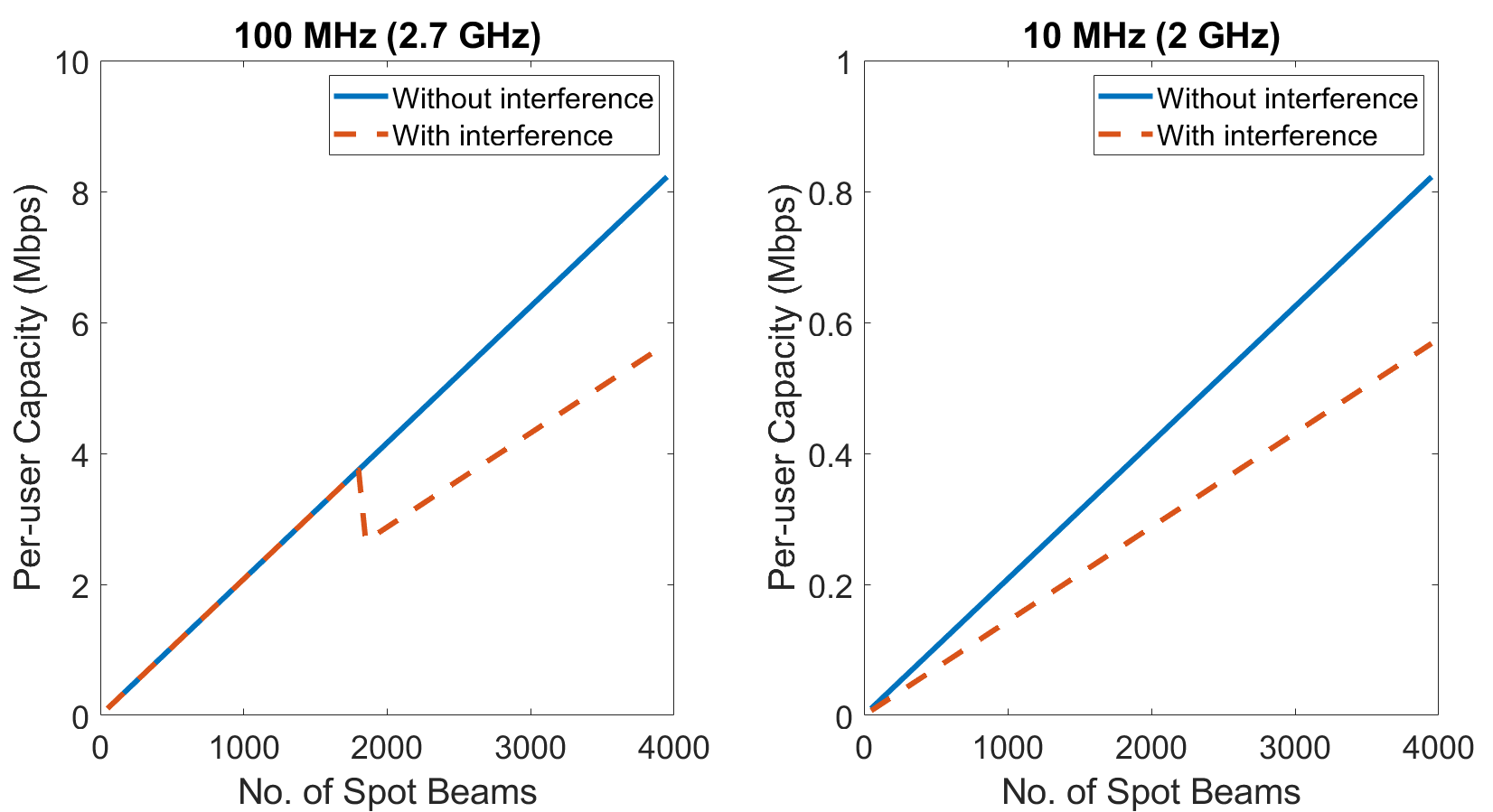}
\caption{The per-user capacity performance in different scenarios.}
\vspace{-1.5em}
\label{Res_cap}
\end{figure}

\subsection{Key Results and Insights}
\begin{figure}
\centering
\includegraphics[scale=0.2]{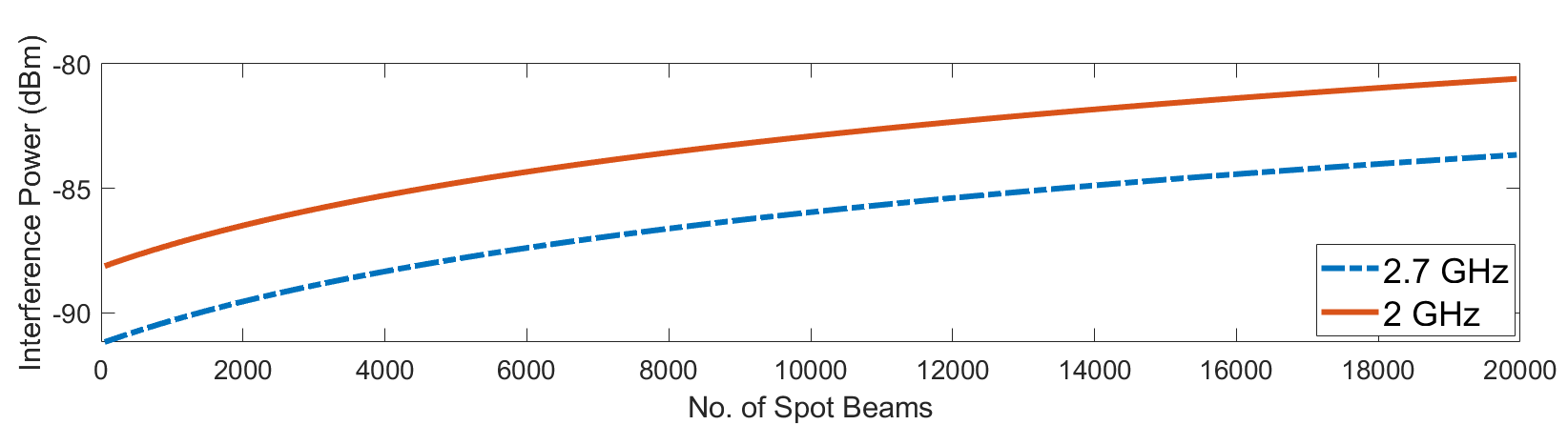}
\caption{Interference power against the number of spot beams.}
\label{Int_beams}
\vspace{-1.5em}
\end{figure}

Fig. \ref{Res_cap} shows the per-user capacity performance against the number of overall spot beams. As expected, the data rate per user increases with spot beams, while interference degrades the capacity per spot beam and user performance. For the 100 MHz scenario, the user capacity decreases and then increases again as the spectral efficiency changes due to the drop in SINR. 
Fig. \ref{Int_beams} illustrates that interference power increases with the number of spot beams, as expected. This trend also confirms the validity of the inter-beam interference modeling. The comparatively higher interference power observed in the 2 GHz band is attributed to its superior propagation characteristics.
\begin{figure}
\centering
\includegraphics[scale=0.2]{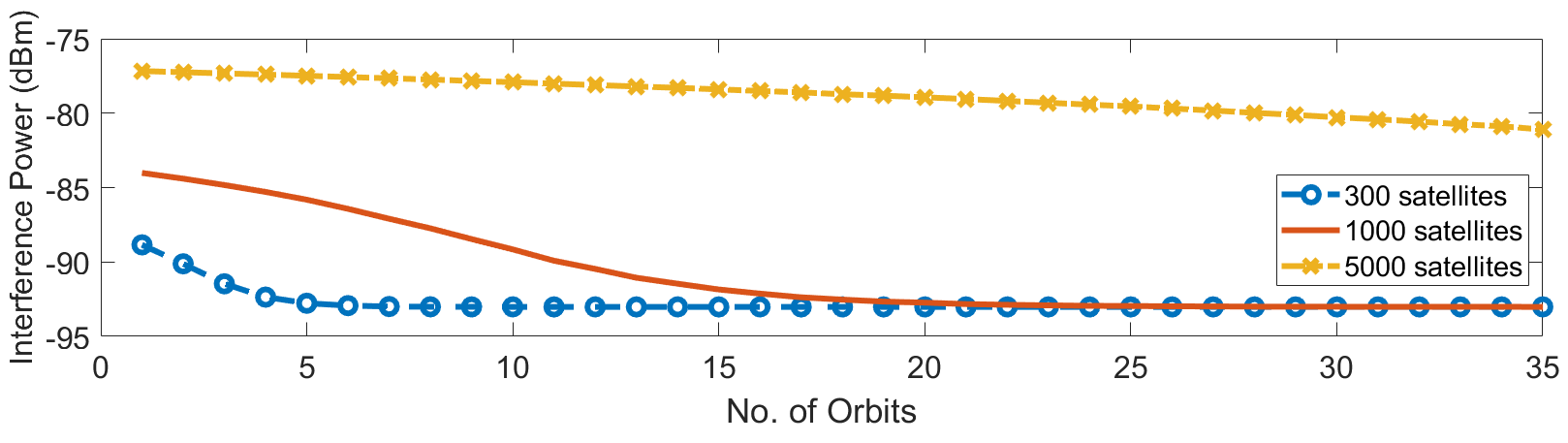}
\caption{Interference power against the number of orbits (100 MHz).}
\label{Res_orbits}
\vspace{-1em}
\end{figure}
Fig. \ref{Res_orbits} captures the interference power, with 2800 spot beams per satellite, against the number of orbits such that each orbit has a different inclination. The results show that inter-satellite interference can be reduced by deploying satellites in different orbits. 


Our assessment reveals the following number of required satellites for providing a minimum data rate guarantee of 10 Mbps per user, with an average global population density of 181 users per km\(^{\textrm{2}}\).

\begin{itemize}
    \item System A with 100 MHz - 14663 satellites.
    \item System B with 10 MHz and 2800 spot beams per satellite - 3351 satellites.
    \item System B with 10 MHz and 1000 spot beams per satellite - 9384 satellites.
\end{itemize}

\begin{figure}
\centering
\includegraphics[scale=0.48]{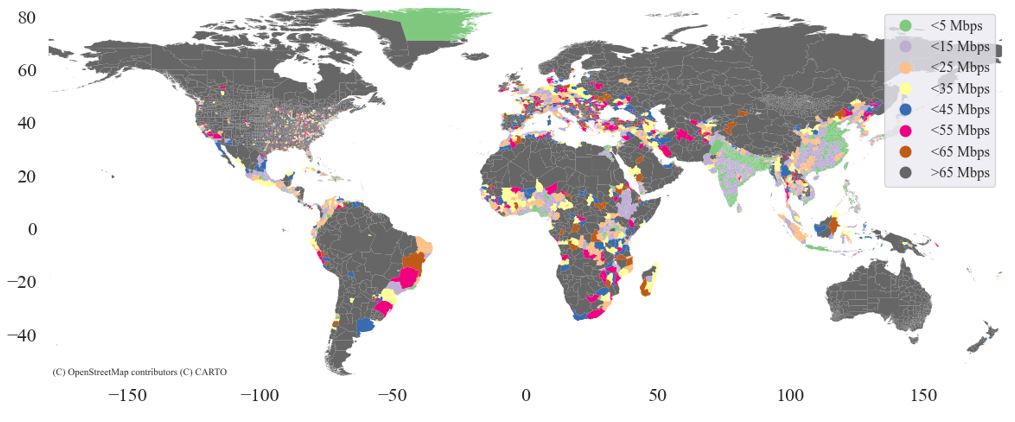}
\caption{Global per-user capacity for System A with 100 MHz bandwidth.}
\label{global_user_capacity_coverage}
\vspace{-1em}
\end{figure}

Fig. \ref{global_user_capacity_coverage} shows the per-user capacity performance in different regions of the world. Note that the user performance varies as per the population density in different regions (Fig. \ref{fig:user_density_distribution}). Similar global capacity performance has been observed for System B with 2800 spot beams per satellite. 

Our economic assessment focuses cost-effectiveness and profitability of a global D2D service. We calculate the total cost of ownership through NPV of both capex and opex components over a 5 year period with a discount rate of 5\%. The NPV is used to estimate the cost per subscriber which in turn determines the net revenue based on the adoption rate.

\begin{figure}
\centering
\includegraphics[scale=0.19]{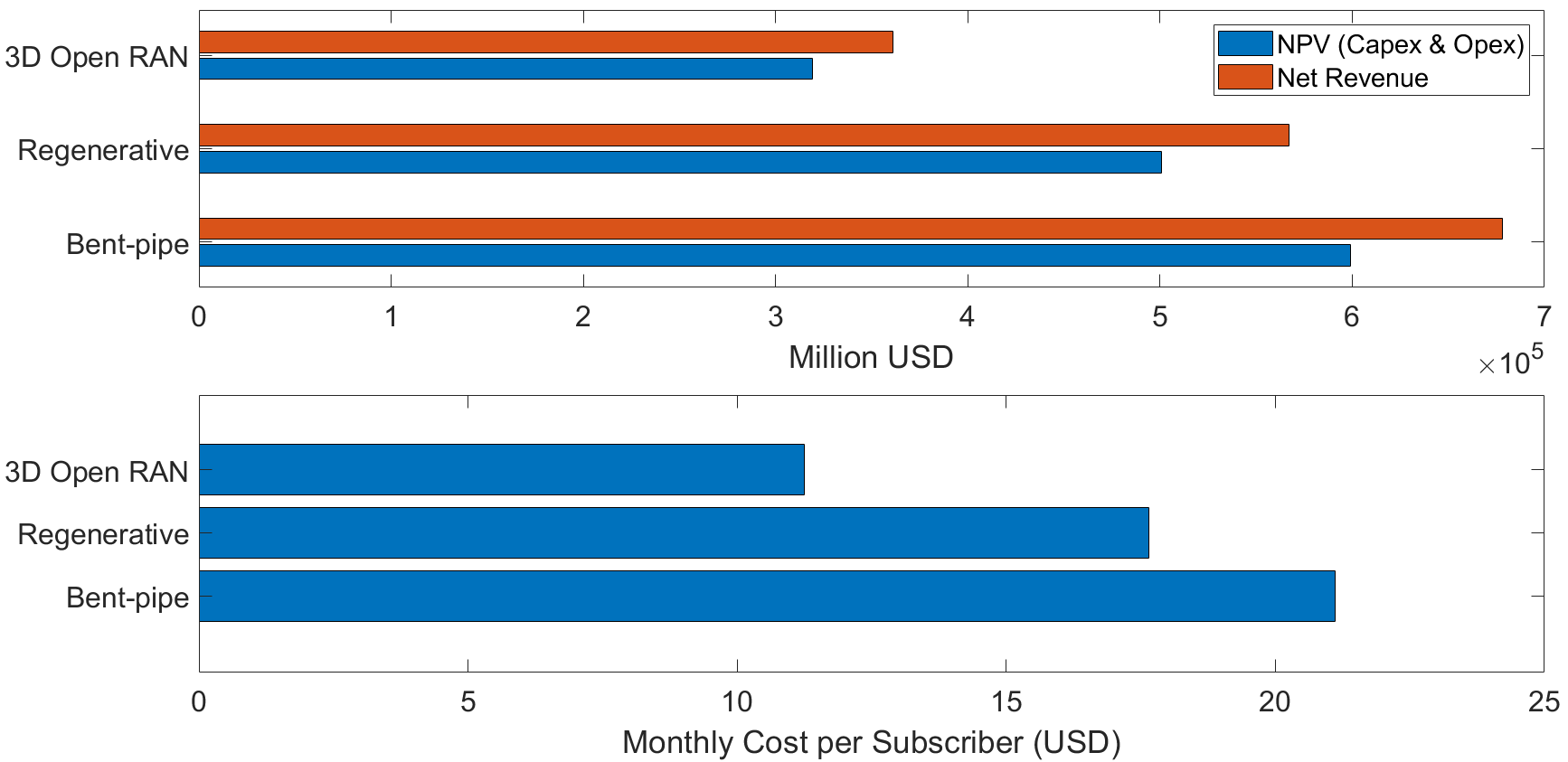}
\caption{Economic assessment for System A (64 beams; 100 MHz).}
\label{Res_TEA_64}
\vspace{-1.5em}
\end{figure}

\begin{figure}
\centering
\includegraphics[scale=0.19]{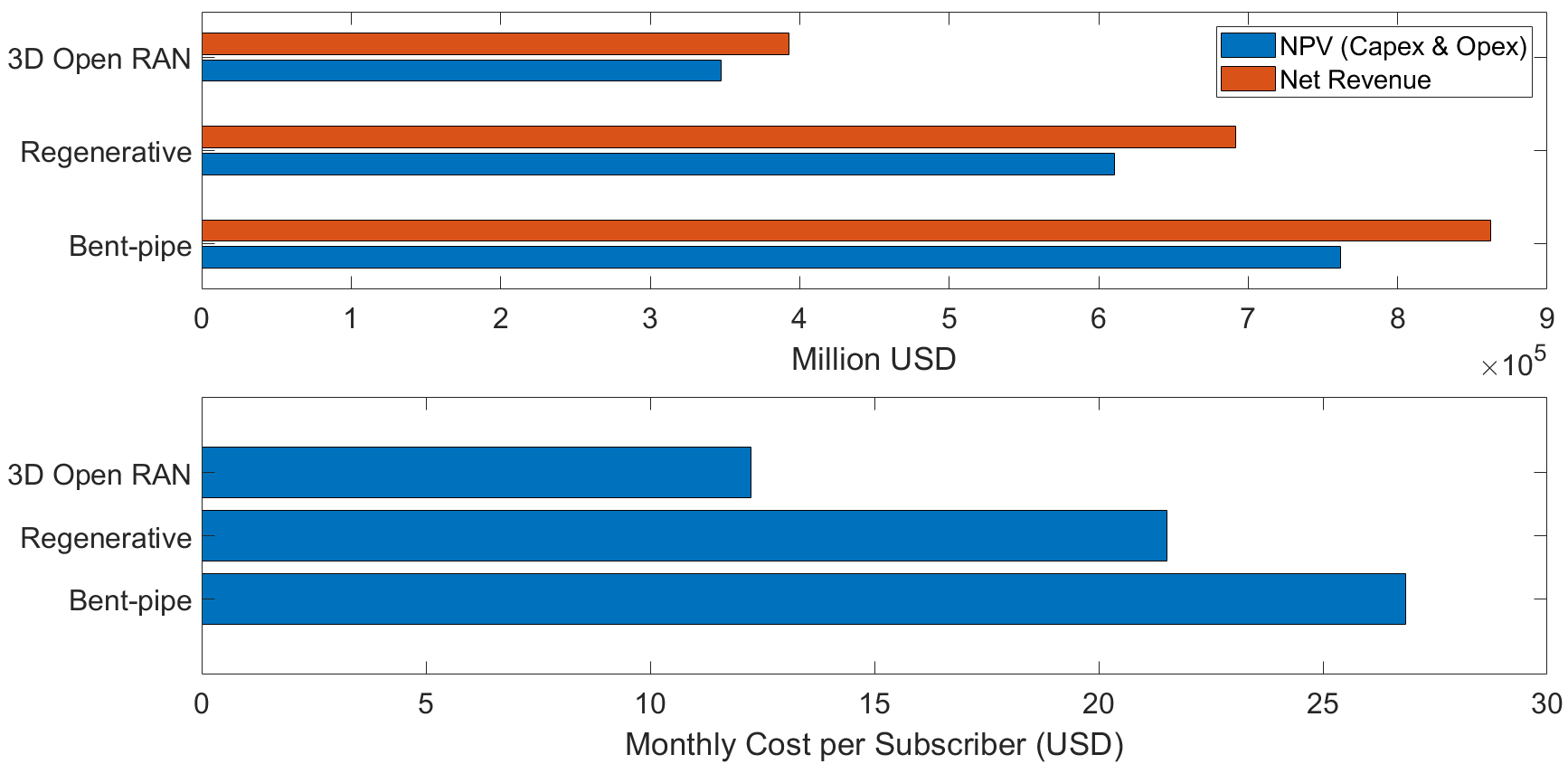}
\caption{Economic assessment for System B (1000 beams; 10 MHz).}
\label{Res_TEA_1000}
\vspace{-1em}
\end{figure}

\begin{figure}
\centering
\includegraphics[scale=0.19]{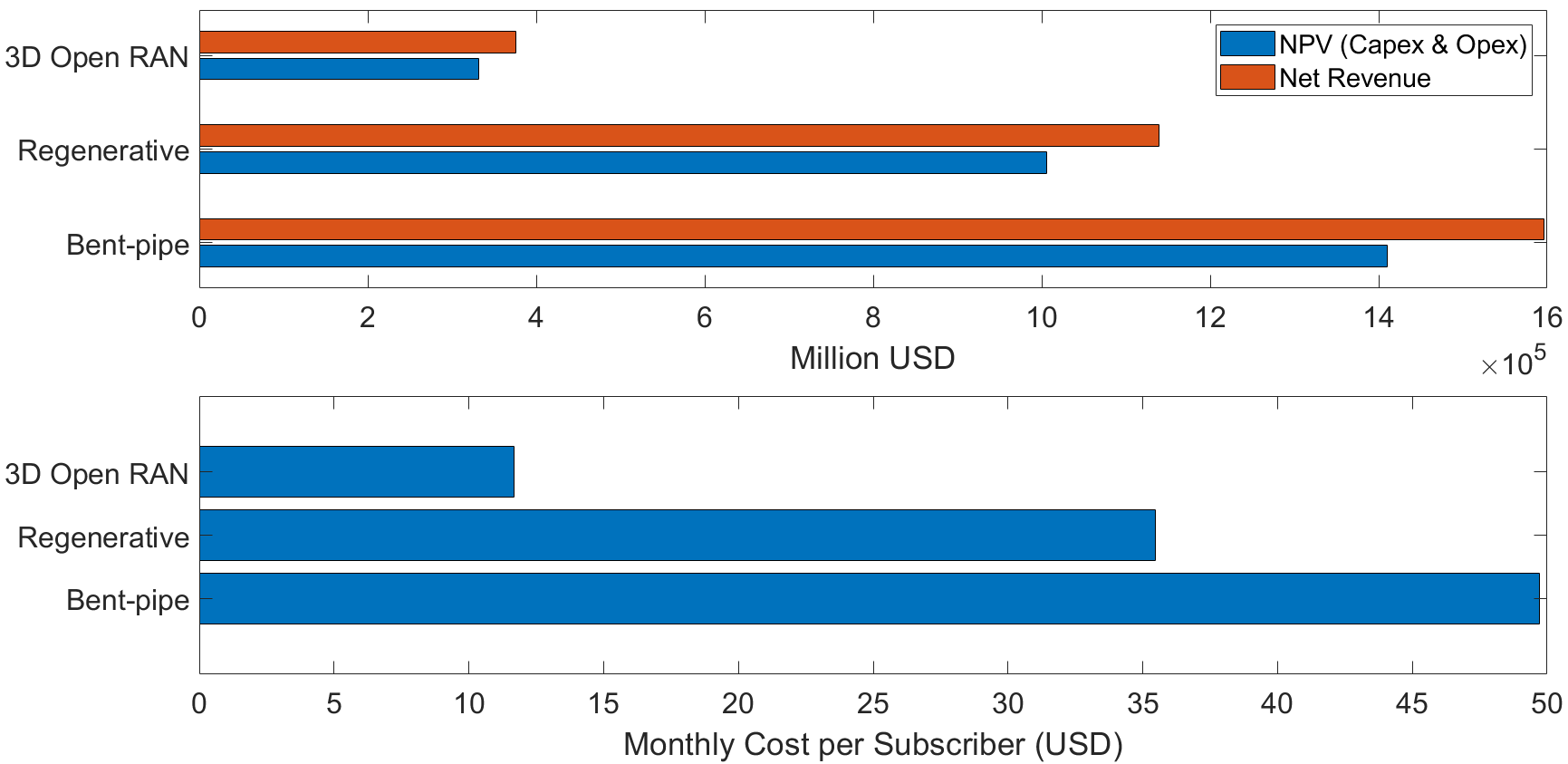}
\caption{Economic assessment for System B (2800 beams; 10 MHz).}
\label{Res_TEA_2800}
\vspace{-1.5em}
\end{figure}

The economic assessment for different systems is shown in Fig. \ref{Res_TEA_64}, Fig. \ref{Res_TEA_1000}, and Fig. \ref{Res_TEA_2800}. Among the three architectural options, the bent-pipe architecture entails the highest cost for building and maintaining a global NTN, which also leads to the highest monthly cost per subscriber for each system. For System A in bent-pipe mode, the monthly cost per subscriber is 21.11 USD. The regenerative payload reduces this cost by 16.4\% (i.e., to 17.65 USD). This reduction  is a result of reduced backhaul and ground station infrastructure costs despite higher costs related to the satellite segment. With 3D Open RAN, this cost reduces by 46.8\% (i.e., from 21.11 to 11.24 USD) and 36.3\% (from 17.65 to 11.24 USD) for bent-pipe and regenerative architectures. This reduction is achieved through disaggregated and low-cost RAN design and reduced ground station infrastructure costs. 

For System B with 1000 spot beams per satellite, the monthly cost per subscriber with bent-pipe, regenerative, and 3D Open RAN architectures is 26.84, 21.52, and 12.24 USD, respectively. With 2800 spot beams per satellite, the respective cost figures are 49.70, 35.46, and 11.68 USD. In the former scenario, the regenerative and 3D Open RAN architectures reduce the cost by 19.8\% and 54.4\%, respectively, as compared to the bent-pipe architecture. In the latter scenario, the cost reduction of regenerative and 3D Open RAN architectures is 40.2\% and 76.5\%, respectively.

System B’s higher cost over System A stems mainly from greater satellite segment expenses, despite needing fewer satellites for global coverage. With 3D Open RAN, however, its monthly cost per subscriber becomes nearly comparable. The 2800-beam configuration is costlier than the 1000-beam setup due to increased satellite platform and launch costs.

\begin{figure}
\centering
\includegraphics[scale=0.19]{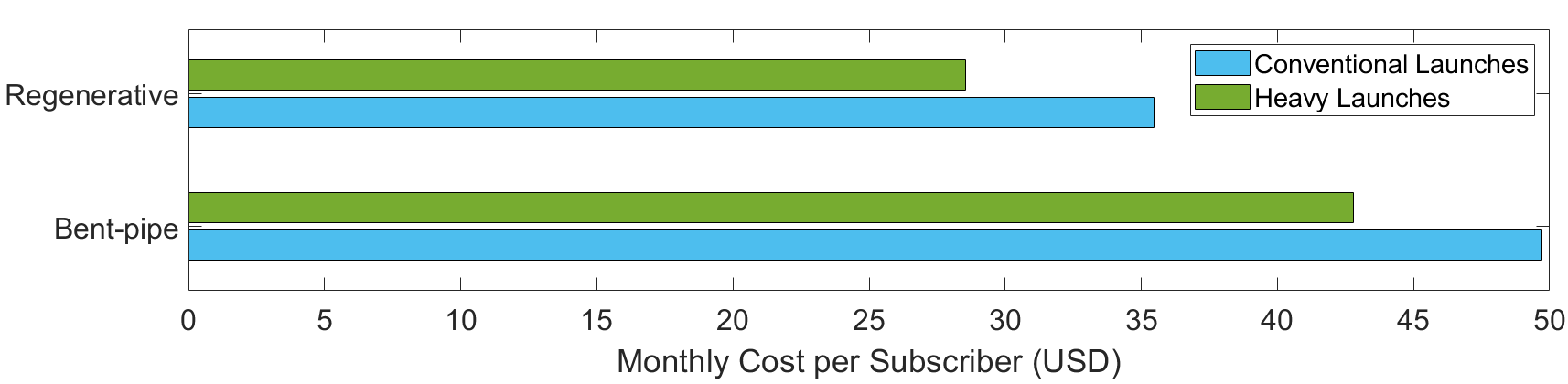}
\caption{Impact of heavy launches on monthly cost per subscriber (System B; 2800 beams; 10 MHz).}
\label{Res_TEA_2800_h}
\vspace{-1.5em}
\end{figure}

Advancements in commercial launch systems, particularly the high payload capacity of vehicles like Starship, have the potential to substantially lower satellite launch costs. Publicly available estimates indicate launch costs around USD 200 per kg, with Starship capable of delivering payloads up to 100 metric tons. As illustrated in Fig. \ref{Res_TEA_2800_h}, our analysis demonstrates that such heavy-lift capabilities can significantly reduce the cost per subscriber. For instance, using heavy launches for System B (featuring 2800 beams and 10 MHz bandwidth) leads to a 13.9\% and 19.5\% reduction in monthly subscriber costs under bent-pipe and regenerative architectures, respectively.

The economic assessment indicates that the net revenue for both systems surpasses their total cost of ownership, resulting in a positive ROI, which is 13.23\% for each system. The regenerative and 3D Open RAN architectures have the potential to yield a higher ROI, contingent on the pricing strategy adopted for the D2D service. 

\section{Concluding Remarks}\label{sect_cr}

This paper introduced a comprehensive TEA framework for global NTN-based satellite D2D, integrating detailed system modeling and cost analysis to assess technical and economic feasibility. It evaluated bent-pipe, regenerative, and 3D Open RAN architectures, highlighting trade-offs between complexity and performance.
We assessed two distinct satellite system architectures designed to deliver a minimum 10 Mbps data rate per user with an average global user density of 181 users per km\(^{\textrm{2}}\). The analysis shows that satellites with limited beamforming capabilities and lower design complexity can offer a more cost-effective solution, particularly when the NTN benefits from greater bandwidth availability (e.g., 100 MHz in MSS spectrum). Conversely, satellites with advanced beamforming and more complex designs can reduce the number of satellites required, but may prove less economical due to the higher costs of the space segment—especially when operating under constrained bandwidth conditions (e.g., 10 MHz leased from a terrestrial MNO). The estimated monthly cost for both systems remains comparable to current terrestrial mobile data offerings (e.g., under 50 USD per user per month). Notably, the adoption of 3D Open RAN technology can reduce the cost per subscriber by more than 50\% to approximately 12 USD per month. Both system designs yield a positive ROI of 13.23\%.
These findings have the potential to inform policy decisions surrounding the deployment and regulation of satellite-based D2D services.

\section*{Acknowledgment}
\vspace{-1.00mm}
This work is partly funded by the UK DSIT as part of the TUDOR project. The authors would  like to thank 
Prof. Ning Wang, Prof. Rahim Tafazolli, and Prof. Barry Evans from the University of Surrey for fruitful discussions.



\bibliographystyle{IEEEtran}

\bibliography{IEEEabrv,TEA}

\end{document}